\documentclass[article,aps,epsfig,float,amsfonts,amssymb,amsmath,shapepar,showkeys,nofootinbib]{revtex4}
\usepackage{graphicx}
\usepackage{dcolumn}
\usepackage{bm}

\usepackage{fancyhdr}
\usepackage{amsmath}
\usepackage{graphicx}
\usepackage{epsfig}
\usepackage{tabularx}
\usepackage{amsmath}
\usepackage{amsthm}
\usepackage{amsfonts}
\usepackage{amssymb}
\usepackage{endnotes}
\usepackage{colortbl}
\usepackage{makeidx}
\usepackage{bm}
\usepackage{bbold}
\usepackage{bbm}
\usepackage[samesize]{cancel}
\usepackage{mathtools}
\usepackage{color}
\usepackage{enumerate}

\begin{document}

\title{Cauchy's formulas for random walks in bounded domains}

\author{Alain Mazzolo}
\email{alain.mazzolo@cea.fr}
\thanks{FAX: +33-1-6908-9490}
\affiliation{CEA/Saclay, DEN/DANS/DM2S/SERMA/LTSD, 91191 Gif-sur-Yvette, France}
\author{Cl\'elia De Mulatier}
\email{clelia.demulatier@cea.fr}
\affiliation{CEA/Saclay, DEN/DANS/DM2S/SERMA/LTSD, 91191 Gif-sur-Yvette, France}
\affiliation{CNRS - Universit\'e Paris-Sud, LPTMS, UMR8626, 91405 Orsay Cedex, France}
\author{Andrea Zoia}
\email{andrea.zoia@cea.fr}
\affiliation{CEA/Saclay, DEN/DANS/DM2S/SERMA/LTSD, 91191 Gif-sur-Yvette, France}

\date{\today}

\begin{abstract}
Cauchy's formula was originally established for random straight paths crossing a body $B \subset \mathbb{R}^{n}$ and basically relates the average chord length through $B$ to the ratio between the volume and the surface of the body itself. The original statement was later extended in the context of transport theory so as to cover the stochastic paths of Pearson random walks with exponentially distributed flight lengths traversing a bounded domain. Some heuristic arguments suggest that Cauchy's formula may also hold true for Pearson random walks with arbitrarily distributed flight lengths. For such a broad class of stochastic processes, we rigorously derive a generalized Cauchy's formula for the average length travelled by the walkers in the body, and show that this quantity depends indeed only on the ratio between the volume and the surface, provided that some constraints are imposed on the entrance step of the walker in $B$. Similar results are obtained also for the average number of collisions performed by the walker in $B$, and an extension to absorbing media is discussed.

\end{abstract}



\maketitle

\section{Introduction}
\label{sec_intro}

Consider a set $M$ of random straight lines in $\mathbb{R}^{d}$ and their intersections $M \cap B$ with an object $B$~\cite{Santalo, Solomon}. These intersections define an ensemble of chords $c$ through $B$: if the measure of random lines is normalized (so that it is a probability measure), then this set of chords can be seen as a random variable, whose associated probability density is usually named chord length distribution (if the object is non-convex, a line may generate several chords, and in this case each chord contributes to the chord length distribution). Furthermore, if the random straight lines obey a uniform density (in a sense that will be made rigorous in Sec.~\ref{sec_length}), then the mean chord length $\langle c \rangle$ is simply related to the volume $V$ and to the surface $\Sigma$ of the body $B$ by~\cite{Santalo}
\begin{equation}
\langle c \rangle = \eta_d \frac{V}{\Sigma},
\end{equation}
where $\eta_d$ is a constant depending only on the dimension $d$ ($\eta_2 = \pi$ and $\eta_3 = 4$). This fundamental result goes under the name of Cauchy's formula. The mean chord length $\langle c \rangle$ is a rather intuitive quantity describing the `mean size' of an object and plays an important role in stereology~\cite{Underwood} and image analysis~\cite{Serra}, since probing structures with random lines allows characterizing a given material. The mean chord length has also become a standard tool in transport theory and nuclear reactor analysis, since the pioneering work of Dirac and co-workers~\cite{Dirac}, especially in the context of external irradiation~\cite{Case}. Moreover, Cauchy's formula has practical applications in various fields such as acoustics~\cite{Kingman}, ecology~\cite{McIntyre}, and is key to the study of random media~\cite{Torquato}. In 1981, Bardsley and Dubi~\cite{Bardsley} based on the linear Boltzmann equation showed that Cauchy's formula applies more generally to the stochastic paths of Pearson random walks~\cite{Pearson} with exponentially distributed flight lengths: for such Markovian processes, the average length $\langle l \rangle$ travelled in a bounded body $B$ is again (and surprisingly) given by $\langle l \rangle = \eta_d V/\Sigma$, when the walkers start uniformly and isotropically on the surface of $B$. This means in particular that the travelled length for exponential Pearson walks is somewhat universal, in that $\langle l \rangle$ is expressed only in terms of the volume and the surface of the traversed medium, independent of the specific details of the geometry and of the walk (such as the mean free path, for instance). This strong result unfortunately went almost unnoticed, until it was independently rediscovered (by following a different approach) by Blanco and Fournier in 2003~\cite{Blanco}, in the context of theoretical ecology (the study of animal displacements through confined geometries). Since then, Cauchy's original statement has attracted intense research efforts, mostly aimed at extending the range of applicability of the formula. The quantity $\langle l \rangle$ was first related to the average length travelled by walkers starting uniformly within the body $B$~\cite{MazzoloEuro}, and successively to the average number of collisions performed in $B$~\cite{Mazzolo05}. B\'enichou and co-workers, based on a backward Chapman-Kolmogorov equation, extended the formula to the higher moments of the travelled length (in relation with the problem of residence times of stochastic motions in bounded domains), explored the effects of mixed boundary conditions (Dirichlet and Neumann) at the surface $\Sigma$, and showed that remarkable ergodic-like properties apply to the average travelled length~\cite{Benichou}. Recently, these results were further generalized by resorting to the Feynman-Kac path-integral approach, which allows explicit formulas to be derived for the travelled lengths and the number of collisions of exponential Pearson walks in the presence of both absorption and branching (this latter representing, e.g., fission events in reactor physics, or the birth of new individuals in the spreading of bacteria as well as of pathogens in epidemics)~\cite{Zoia3}.

Such findings show that Cauchy's formula and exponential Pearson random walks are intimately connected. However, it is well-known that many real-world stochastic transport phenomena do not display exponentially distributed jump lengths (see, e.g., the discussion in~\cite{Zoia2, Zoia6, Zoia4}), and one might naturally wonder whether similar universal properties still hold for these processes. Actually, some heuristic findings suggest that Cauchy's formula applies to arbitrary (isotropic) random walks whenever the random walk enters the domain $B$ with a length distribution compatible with equilibrium~\cite{Benichou, BlancoPRL}. In particular, this was formally proven (by resorting to integral geometry arguments) for the quite restrictive class of walks having a single collision in $B$ before leaving the domain~\cite{Mazzolo09}.

The purpose of this paper is to show under which conditions Cauchy's formula is valid for arbitrary Pearson walks. We begin our analysis by deriving a key result for the average number of collisions in $B$, within the framework of transport theory. Then, by building upon this result and by using the standard tools of geometric probabilities, we proceed further to derive a Cauchy's formula for the average lengths of the random paths. We conclude by considering an extension of these results to the case of purely absorbing random walks.

\section{A Cauchy's formula for the average number of collisions}
\label{sec_N}

In a series of recent papers~\cite{Zoia1, Zoia5, Zoia3}, it has been shown that the travelled length and collision statistics of exponential Pearson walks in bounded domains can be straightforwardly determined by resorting to the Feynman-Kac path-integral formalism. In particular, this approach allows deriving generalized Cauchy's formulas for the moments of any order of lengths and collisions, and accounts also for the possibility of birth-death events leading to branching paths. However, the key argument underlying the applicability of the path-integral approach (at least for the travelled length, in the form discussed in the references above) is the Markovian nature of exponential Pearson walks, and this property is unfortunately not preserved when considering general Pearson walks with arbitrarily distributed jumps. While it is in principle possible to extend the path-integral approach to non-Markovian continuous-time processes (see, e.g.,~\cite{Turgeman} for the case of anomalous diffusion with power-law memory kernels), we will not pursue this idea here. In this paper, we will consider instead a simpler strategy, and resort to standard transport theory. At the same time, we initially restrict our attention to diffusive walks (in other words, we neglect birth-death events).

Consider the random walk of a particle from a point-source ${\cal S}= \delta({\mathbf r}-{\mathbf r_0}) \delta(\boldsymbol{\omega} -\boldsymbol{\omega}_0 )$ located at ${\mathbf r}_0$ on the surface of $B$. The particle is emitted in the inward direction ${\boldsymbol{\omega}}_0$, and moves at constant speed. The stochastic trajectory is composed of a series of flights of random length, interrupted by collisions (whereupon the particle is scattered, i.e., randomly changes its direction $\boldsymbol{\omega}$), and terminates when the particle crosses the surface of $B$ for the first time. This formally corresponds to imposing Dirichlet (or `vacuum') boundary conditions on the surface of $B$: a particle crossing the boundary is considered to be lost and can not re-enter the domain. In order to fully characterize collisions, we introduce the density $C({\boldsymbol{\omega}}|{\boldsymbol{\omega}'},{\mathbf r})$, namely, the conditional probability density of changing direction from ${\boldsymbol{\omega}'}$ to ${\boldsymbol{\omega}}$, given a scattering event at ${\mathbf r}$. The displacement law for the first jump (which characterizes the particle entrance in $B$, from the starting point ${\mathbf r_0}$ and in direction ${\boldsymbol{\omega}}_0$) is denoted $H({\mathbf r}|{\mathbf r_0},{\boldsymbol{\omega}_0})$. All other jump lengths following a collision at a point ${\mathbf r'}$ in $B$ (if any) obey the density $T({\mathbf r}|{\mathbf r'},{\boldsymbol{\omega}'})$, where ${\boldsymbol{\omega}'}$ is the current direction of the walker. We enumerate successive collisions by a discrete index $g$.

It follows that the quantity
\begin{equation}
K({\mathbf r},{\boldsymbol{\omega}}|{\mathbf r'},{\boldsymbol{\omega}'}) = T({\mathbf r}|{\mathbf r'},{\boldsymbol{\omega}})C({\boldsymbol{\omega}}|{\boldsymbol{\omega}'},{\mathbf r'})
\end{equation}
represents the density of particles entering a collision at the $(g+1)$-th generation with coordinates $\left\lbrace {\mathbf r},{\boldsymbol{\omega}} \right\rbrace$, having entered a collision at the $g$-th generation with coordinates $\left\lbrace {\mathbf r'},{\boldsymbol{\omega}'} \right\rbrace$. We furthermore introduce the incident particle density $\Psi_g({\mathbf r},{\boldsymbol{\omega}}|{\mathbf r_0},{\boldsymbol{\omega}_0})$, which is defined such that $N_{A} = \int_A d{\mathbf r} \int_{\Omega_d} d{\boldsymbol{\omega}} \Psi_g({\mathbf r},{\boldsymbol{\omega}}|{\mathbf r_0},{\boldsymbol{\omega}_0})$ is the average number of collisions in the region $A$ at the $g$-th generation. Then, the stationary incident particle density will be given by
\begin{equation}
\Psi({\mathbf r},{\boldsymbol{\omega}}|{\mathbf r_0},{\boldsymbol{\omega}_0})=\lim_{N \to \infty} \sum_{g=1}^N \Psi_g({\mathbf r},{\boldsymbol{\omega}}|{\mathbf r_0},{\boldsymbol{\omega}_0}),
\end{equation}
provided that such limit exists~\cite{Spanier, Lux}. As customary, the source contribution is not taken into account in the definition of $\Psi({\mathbf r},{\boldsymbol{\omega}}|{\mathbf r_0},{\boldsymbol{\omega}_0})$~\cite{Spanier}. It can be shown that the stationary incident particle density satisfies the linear integral transport equation~\cite{Lux, Zoia5}
\begin{equation}
\Psi({\mathbf r},{\boldsymbol{\omega}}|{\mathbf r_0},{\boldsymbol{\omega}_0})= \int d{\mathbf r}' \int d\boldsymbol{\omega}' K({\mathbf r},{\boldsymbol{\omega}}|{\mathbf r'},{\boldsymbol{\omega}'})  \Psi({\mathbf r}', \boldsymbol{\omega}'|{\mathbf r_0},{\boldsymbol{\omega}_0})+ H({\mathbf r}|{\mathbf r}_0,{\boldsymbol{\omega}_0}) \delta(\boldsymbol{\omega} -\boldsymbol{\omega}_0 ) ,
\label{integral_transport_equations_one}
\end{equation}
with Dirichlet boundary conditions $\Psi({\mathbf r},\boldsymbol{\omega}|{\mathbf r_0},{\boldsymbol{\omega}_0}) = 0$ for $\boldsymbol{\omega}_0$ directed outwards. Generally speaking, closed-form solutions to Eq.~\eqref{integral_transport_equations_one} can hardly be found for arbitrary geometries~\cite{Wasow}. However, our aim is now to average Eq.~\eqref{integral_transport_equations_one} over an appropriate source condition at the boundary. Having in mind the case of straight paths and of exponential Pearson walks, we choose here an isotropic and uniform particle distribution on the surface of $B$ (a $\mu-$ randomness in the language of stochastic geometry, which is (up to a constant) the unique measure invariant under the group of motion~\cite{Santalo}). This choice allows the case of straight paths to be recovered when the density $H$ is such that the first jumps are much larger than the domain size. In this case, the probability measure at the surface reads~\cite{Benichou, Zoia3}
\begin{equation}
\eta_d \frac{d\Sigma}{\Sigma} \frac{d\boldsymbol{\omega}_0}{\Omega_d} (\boldsymbol{\omega}_0 \cdot \bf n),
\end{equation}
where $\Omega_d=2\pi^{d/2}/\Gamma(d/2)$ is the surface of the unit sphere in dimension $d$ and
\begin{equation}
\eta_d = \sqrt{\pi}(d-1)\frac{\Gamma \left(\frac{d-1}{2} \right)}{\Gamma \left( \frac{d}{2} \right)}
\end{equation}
is a dimension-dependent normalization constant, equal to twice the inverse of the average height of the $d$-dimensional unit shell, $\boldsymbol{n}$ being the unit vector normal to the surface and pointing inwards (remark that in~\cite{Benichou} the normal was taken to be pointing outwards). We can then properly define the average of $\Psi({\mathbf r},{\boldsymbol{\omega}}|{\mathbf r_0},{\boldsymbol{\omega}_0})$ over the surface of $B$ by taking
\begin{equation}
\langle \Psi \rangle_\Sigma({\mathbf r},{\boldsymbol{\omega}}) = \eta_d \int \frac{d\Sigma }{\Sigma} \int \frac{d\boldsymbol{\omega}_0}{\Omega_d} (\boldsymbol{\omega}_0   \cdot \mathbf{n}) \Psi({\mathbf r},{\boldsymbol{\omega}}|{\mathbf r_0},{\boldsymbol{\omega}_0}).
\label{eq.mean_surf}
\end{equation}
In the following, we will furthermore assume that scattering is isotropic, which implies
\begin{equation}
C({\boldsymbol{\omega}}|{\boldsymbol{\omega}'},{\mathbf r'}) =  \frac{1}{\Omega_d}.
\end{equation}
Under this hypothesis, taking the surface average of Eq.~\eqref{integral_transport_equations_one} leads to
\begin{align}
\eta_d \int \frac{d\Sigma }{\Sigma} \int \frac{d\boldsymbol{\omega}_0}{\Omega_d} (\boldsymbol{\omega}_0 \cdot \mathbf{n}) \Psi({\mathbf r},{\boldsymbol{\omega}}|{\mathbf r_0},{\boldsymbol{\omega}_0}) & =\int d{\mathbf r}' \, T({\mathbf r}|{\mathbf r'},{\boldsymbol{\omega}}) \int \frac{d\boldsymbol{\omega}'}{\Omega_d}\left[ \eta_d \int \frac{d\Sigma }{\Sigma} \int \frac{d\boldsymbol{\omega}_0}{\Omega_d} (\boldsymbol{\omega}_0   \cdot \mathbf{n}) \Psi({\mathbf r'},{\boldsymbol{\omega}'}|{\mathbf r_0},{\boldsymbol{\omega}_0})   \right] \notag \\
   & + \eta_d \int \frac{d\Sigma }{\Sigma} \int \frac{d\boldsymbol{\omega}_0}{\Omega_d} (\boldsymbol{\omega}_0 \cdot \mathbf{n})
         H({\mathbf r}|{\mathbf r}_0,{\boldsymbol{\omega}_0}) \delta(\boldsymbol{\omega} -\boldsymbol{\omega}_0 ).
\end{align}
Then, by making use of the definition of the surface average given in Eq.~\eqref{eq.mean_surf}, we can rewrite
\begin{equation}
\langle \Psi \rangle_\Sigma({\mathbf r},{\boldsymbol{\omega}}) = \int d{\mathbf r}' \int \frac{d\boldsymbol{\omega}'}{\Omega_d} \, T({\mathbf r}|{\mathbf r'},{\boldsymbol{\omega}}) \langle \Psi \rangle_\Sigma({\mathbf r'},{\boldsymbol{\omega}'}) + \frac{\eta_d} {\Omega_d \Sigma}  \int d\Sigma (\boldsymbol{\omega} \cdot \mathbf{n}) H({\mathbf r}|{\mathbf r}_0,{\boldsymbol{\omega}}).
\label{eq.psi.average}
\end{equation}

\begin{figure}[!h]
\centering
\includegraphics[width=3.in,height=2.2in]{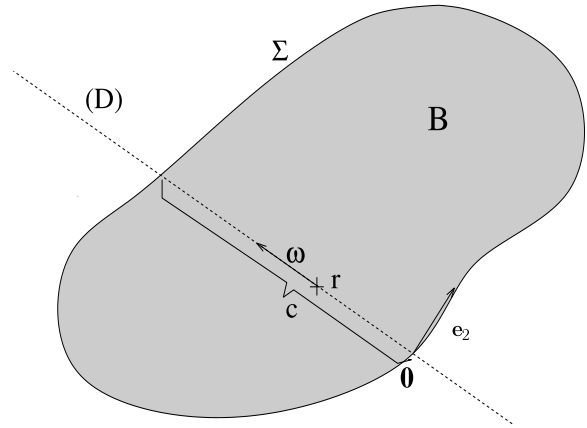}
\setlength{\abovecaptionskip}{15pt} 
\caption{Two-dimensional projection in the frame $(0,{\boldsymbol{\omega}},\mathbf e_2)$.}
\label{fig1}
\end{figure}

We now ask whether there exists a first-jump distribution $H$ such that the resulting incident collision density $\langle \Psi \rangle_\Sigma({\mathbf r},{\boldsymbol{\omega}})$ in Eq.~\eqref{eq.psi.average} is uniform in $B$, i.e., $\langle \Psi \rangle_\Sigma({\mathbf r},{\boldsymbol{\omega}}) = \langle \Psi \rangle_\Sigma$, which intuitively corresponds to an equilibrium condition for the particles (see, e.g., the discussion in~\cite{BlancoPRL}). If this is the case, we obtain
\begin{equation}
\langle \Psi \rangle_\Sigma - \langle \Psi \rangle_\Sigma\int d{\mathbf r}' \, T({\mathbf r}|{\mathbf r}',{\boldsymbol{\omega}}) = \frac{\eta_d} {\Omega_d \Sigma} \int d\Sigma (\boldsymbol{\omega} \cdot \mathbf{n}) H({\mathbf r}|{\mathbf r}_0,{\boldsymbol{\omega}}).
\label{eq.psi.average.aux}
\end{equation}
By applying the divergence theorem at the r.~h.~s.~of Eq.~\eqref{eq.psi.average.aux} we get then
\begin{equation}
\frac{\Omega_d \Sigma}{\eta_d} \langle \Psi \rangle_\Sigma \left[1- \int d{\mathbf r}' \, T({\mathbf r}|{\mathbf r}',{\boldsymbol{\omega}}) \right] = - \int d{\mathbf r}_0 \nabla \cdot \left[ \boldsymbol{\omega} H({\mathbf r}|{\mathbf r}_0,{\boldsymbol{\omega}}) \right].
\label{eq.psi.average.aux2}
\end{equation}
Observe that the integrals in Eq.~\eqref{eq.psi.average.aux} are over the volume of $B$, whereas the jump densities $ T({\mathbf r}|{\mathbf r'},{\boldsymbol{\omega}})$ and $H({\mathbf r}|{\mathbf r}_0,{\boldsymbol{\omega}})$ are non-zero only when $\mathbf r'-\mathbf r$ and $\mathbf r - \mathbf r_0$, respectively, are aligned with the direction $\boldsymbol{\omega}$ (this is illustrated in Fig.~\ref{fig1}). We can therefore make use of this property to perform the integration on an appropriate basis.

Let us define $(D)$ the line passing by $\mathbf r$ in direction $\boldsymbol{\omega}$, and denote $\mathbf 0$ the intersection of $(D)$ with the surface in the direction opposite to $\boldsymbol{\omega}$ (see Fig.~\ref{fig1}). We can then work in the ortho-normal frame $(\mathbf{0},\boldsymbol{\omega}, \mathbf e_2, \ldots, \mathbf e_d)$, where $\mathbf e_2, \ldots, \mathbf e_d$ are a set of $d-1$ vectors orthogonal to $\boldsymbol{\omega}$. In this orthonormal basis, $\mathbf r =  r \boldsymbol{\omega}$, with $r>0$, and the densities $H$ and $T$ take the form
\begin{align}
\label{T_delta}
H({\mathbf r}|{\mathbf r}_0,{\boldsymbol{\omega}}) &= h(r-{r_0}_1) \delta({r_0}_2) \ldots \delta({r_0}_d)\\
T({\mathbf r}|{\mathbf r}',{\boldsymbol{\omega}}) &= t(r-r_1') \delta(r_2') \ldots \delta(r_d') ,\notag
\end{align}
where $h(x)$ is the probability density of the first jump length and $t(x)$ that of the following jump lengths. In the same basis, we have
\begin{equation}
\nabla \cdot \left[ \boldsymbol{\omega} H({\mathbf r}|{\mathbf r}_0,{\boldsymbol{\omega}}) \right] = \partial_{{r_0}_1} h(r-{r_0}_1) \delta({r_0}_2) \ldots \delta({r_0}_d) = - h'(r-{r_0}_1)\delta({r_0}_2) \ldots \delta({r_0}_d).
\label{H_delta}
\end{equation}
Replacing Eqs.~\eqref{T_delta} and~\eqref{H_delta} into Eq.~\eqref{eq.psi.average.aux2} and performing the integrations over delta distributions, we obtain
\begin{equation}
\frac{\Omega_d \Sigma}{\eta_d} \langle \Psi \rangle_\Sigma \left[1- \int_0^c dr_1' \, t(r-r_1') \right] = \int_0^c d{r_0}_1 h'(r-{r_0}_1),
\label{eq.p.and.t}
\end{equation}
where $c$ is the length of the chord resulting from the intersection of $(D)$ with $B$. Since both functions $t(x)$ and $h(x)$ must vanish for $x<0$ (jump lengths being positive), we immediately get
\begin{equation}
\frac{\Omega_d \Sigma}{\eta_d} \langle \Psi \rangle_\Sigma \left[1- \int_0^r dx \, t(x) \right] = h(r).
\end{equation}
Furthermore, $h(r)$ is a probability density, and normalization yields
\begin{equation}
\langle \Psi \rangle_\Sigma = \frac{\eta_d}{\Omega_d \Sigma \lambda},
\label{eq.mean.psi}
\end{equation}
where $\lambda = \int_0^{\infty} dx \, t(x) x$ is the average jump length of an unconstrained walker. Hence, we have the sought condition on $H$, namely,
\begin{equation}
h(r) = \frac{1}{\lambda} \int_r^{\infty} dx \, t(x).
\label{eq.h}
\end{equation}
Observe that we should require $\lambda$ to be finite in order for $h(r)$ to be properly normalized. For instance, if $t(x)$ is a L\'evy-stable density, with $t(x)=x^{-1-\alpha}$ for $x \to \infty$~\cite{levy}, $\lambda$ is not finite for $0 < \alpha \leq 1$ and Eq.~\eqref{eq.h} does not apply.  

Equation~\eqref{eq.h} can be seen a condition relating the first-jump density $H$ to the jump density $T$ in $B$: if this condition is met, then the stationary collision density in $B$ is uniform, i.e., particles are at equilibrium. Moreover, since the mean number of collisions in $B$ is given by the integral over the phase space
\begin{equation}
\langle N \rangle_\Sigma  = \int_V  d{\mathbf r} \int_{\Omega_d} d\boldsymbol{\omega} \langle \Psi \rangle_\Sigma,
\label{phase_space}
\end{equation}
replacing Eq.~\eqref{eq.mean.psi} into Eq.~\eqref{phase_space} leads to
\begin{equation}
\langle N \rangle_\Sigma = \frac{\eta_d}{\lambda} \frac{V}{\Sigma},
\label{eq.mean.N}
\end{equation}
which is a generalization to arbitrary Pearson random walks of the Cauchy's formula derived in~\cite{Mazzolo05} for the average number of collisions of exponential Pearson walks. 

The relevance of the assumption of isotropic scattering can be understood by considering, e.g., random walks having a constant jump length $a$ (i.e., $t(x)=\delta(x-a)$), very small as compared to the domain size. Such walks enter the domain with a jump length that is at most equal to $a$ (see Eq.~\eqref{eq.h}). In this case, the scattering distribution can not be arbitrary: for instance, a systematically back-scattering angular distribution would force particles to immediately leave from the boundaries, so that the domain would not be entirely probed. Then, a reasonable choice is to assume an isotropic distribution. For the particular case of exponentially distributed Pearson walks, the isotropy condition can be actually relaxed by demanding detailed balance to be satisfied at each collision~\cite{Bardsley}.

We might further wonder whether there exists a process for which $h(x)=t(x)$, i.e., the law of the first jump coincides with that of the others. By imposing then $t(r)= (1/\lambda) \int_r^{\infty} dx \, t(x)$, it immediately follows that this requirement is met by exponential Pearson random walks with density $t(x)=h(x)=(1/\lambda)\exp(-x/\lambda)$. This is intuitively due to the Markovian (memoryless) nature of such walks (see, e.g.,~\cite{Blanco, MazzoloEuro, Benichou}).

\section{Average travelled length}
\label{sec_length}

\begin{figure}[h]
\centering
\includegraphics[width=5.8in,height=2.in]{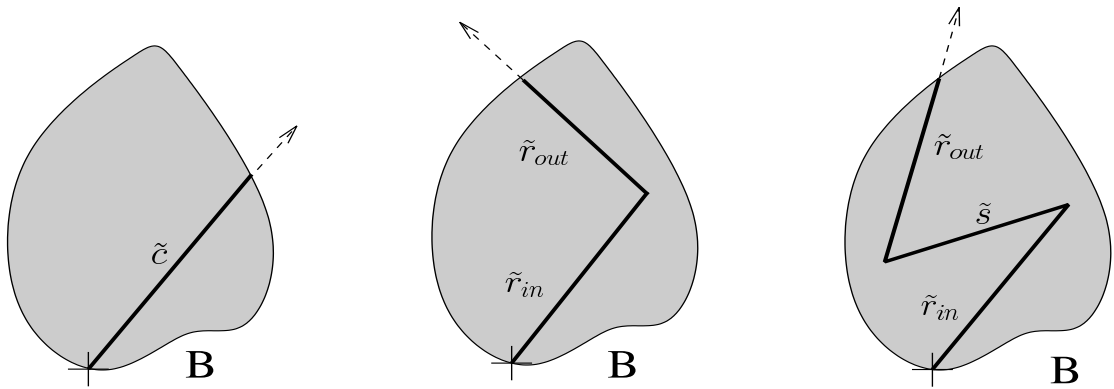}
\setlength{\abovecaptionskip}{15pt}  
\caption{A schematic representation of trajectories having $0,1,2$ collisions inside $B$ and their associated lengths.}
\label{fig2}
\end{figure}

We will now address the issue of determining the average travelled length in $B$. Assume that the same hypotheses as above hold concerning the stochastic process and the distribution of the starting points. Let $P_n$ be the probability that a trajectory entering the domain has exactly $n$ collisions inside $B$ ($P_0$ is thus the probability that the trajectory is a chord, which precisely happens if the stochastic path has no collisions in $B$). Let us furthermore define $\langle\,L_n\,\rangle_\Sigma$ the mean length travelled by paths constrained to have performed a number $n$ of collisions in $B$. Since trajectories having exactly $n$ collisions form a complete set of disjoint events, the mean length $\langle\,L\,\rangle_\Sigma$ of trajectories inside the domain will be given by
\begin{equation}
\label{Lmean_start}
 \langle\,L\,\rangle_\Sigma=\sum_{n=0}^{\infty} P_n \langle\,L_n\,\rangle_\Sigma.
\end{equation}
As the walk is constrained to start from the surface and is terminated at the first exit from the surface, the quantity $\langle\,L\,\rangle_\Sigma$ is directly related to the so-called first-return times to the boundary (the velocity of the particle being assumed constant)~\cite{Blanco}. As shown in Fig.~\ref{fig2}, stochastic trajectories consist of a series of segments which can enter from the surface, lie entirely within $B$, or leave from the surface. In order to fully characterize these distinct configurations, we introduce the following notation:
\begin{itemize}
\renewcommand{\labelitemi}{\textbullet}
\item $\tilde{c}$, a random chord (the path undergoes no collisions inside the domain), with associated density $p_{\tilde{c}}$.
\item $\tilde{r}_{in}$, the first jump length inside $B$ (the starting point being uniformly distributed on the surface), with associated density $p_{in}$.
\item $\tilde{r}_{out}$, the last jump length (the final point lies outside the domain), with associated density $p_{out}$.
\item $\tilde{s}$, a segment entirely contained the domain, with associated density $p_{\tilde{s}}$.
\end{itemize}
The tilde is used here to recall that all these quantities are random variables depending on the realizations of $h(x)$ (for segments starting from the surface) or $t(x)$ (for segments starting inside $B$).

With these notations, we can identify
\begin{align} 
   & \langle\,L_0\,\rangle_\Sigma = \langle\,\tilde{c}\,\rangle                   \notag\\
   & \langle\,L_n\,\rangle_\Sigma = \langle\,\tilde{r}_{in}\,\rangle + (n-1)\langle\,\tilde{s}\,\rangle + \langle\,\tilde{r}_{out}\,\rangle  \mathrm{~~for~~} n \geq 1  \notag.
\end{align}
Replacing these quantities in Eq.~\eqref{Lmean_start} leads then to
\begin{equation}
	\langle\,L\,\rangle_\Sigma = \langle\,\tilde{c}\,\rangle P_0 + (\langle\,\tilde{r}_{in}\,\rangle+\langle\,\tilde{r}_{out}\,\rangle)\sum_{n=1}^{\infty} P_n + \langle\,\tilde{s}\,\rangle \sum_{n=1}^{\infty} (n-1)P_n.
\end{equation}
From normalization, $\sum_{n=0}^{\infty} P_n=1$, and observe that by definition $\langle\,N\,\rangle_\Sigma=\sum_{n=1}^{\infty} n P_n$. It follows that
\begin{equation}
\label{L_start}
 \langle\,L\,\rangle_\Sigma =  \langle\,\tilde{c}\,\rangle P_0 + (\langle\,\tilde{r}_{in}\,\rangle+\langle\,\tilde{r}_{out}\,\rangle )(1- P_0) + \langle\,\tilde{s}\,\rangle \left[\langle\,N\,\rangle_\Sigma -(1-P_0)\right]  .
\end{equation}
We have now to explicitly compute each term appearing in this equation. The term $\langle\,N\,\rangle_\Sigma=\langle\,c\,\rangle/\lambda$, where $\langle\,c\,\rangle=\eta_d V/S$, is known from the results derived in the previous Section.

\subsection{The term $P_0$}
\label{sec_P_0}

We begin by considering the term $P_0$, which is the probability that the final point of the first jump actually falls outside $B$, i.e., that the length $r$ of the first jump is larger than the supported chord. (The chord length distribution is defined as  $\mathrm {F(c) = Prob\{c(M) \leq c:M \cap B \not= \varnothing \} }$, measured with the uniform density $\mathrm{M}$ of random lines in the sense of the theory of geometric probability (the $\mu-$ randomness)~\cite{Santalo,Solomon}, and $f(c)= dF(c)/dc$ is  the corresponding density function. This assumptions is sometimes denoted IUR-chords (Isotropic Uniform Random chords), since this randomness stems from the body being exposed to a uniform, isotropic field of straight infinite
lines~\cite{Kellerer}). From this definition we have
\begin{equation}
 \displaystyle P_0 = \mathrm{Prob[} c \le r \mathrm{]} = \int_0^{\infty} dr \, h(r) \int_0^{r} dc \, f(c) =  \int_0^{\infty} dc \, f(c) \int_{c}^{\infty} dr \, h(r) .
\end{equation}
From Eq.~\eqref{eq.h} we can rewrite
\begin{align}
\int_{c}^{\infty}  dr \, h(r) & = \frac{1}{\lambda} \int_{c}^{\infty} dr \int_r^{\infty} dx \, t(x) =  \frac{1}{\lambda} \int_c^{\infty} dx \, t(x) \int_{c}^{x} dr =  \frac{1}{\lambda} \int_c^{\infty} dx \, t(x) (x-c) \notag\\
							& = 1 -\frac{c}{\lambda} + \frac{1}{\lambda} \int_0^{c} dx \, t(x) (c-x).
\label{eq_g(r)}
\end{align}
Then, replacing this expression in the definition of $P_0$ yields (after having permuted the order of integration)
\begin{equation}
  \label{P_0_final}
	P_0 = \frac{1}{\lambda} \int_0^{\infty} dx \, t(x) \int_0^{x} dc \, f(c)(x-c).
\end{equation}

\begin{figure}[!h]
\centering
\includegraphics[width=1.8in,height=1.4in]{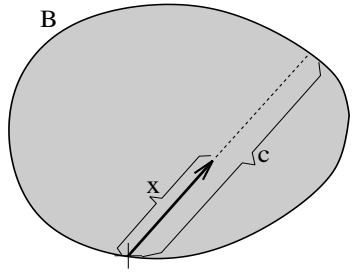}
\setlength{\abovecaptionskip}{15pt} 
\caption{First jump lengths inside $B$ are smaller than supported chords.}
\label{fig3}
\end{figure}

\subsection{Proof that $p_{in} = p_{out}$.}

The random path begins with a jump from the surface whose length (independent of whether the jump falls or not inside $B$) obeys the density $h(x)$. Thus, the probability density $p_{in}(x)$ of the first jump inside $B$ is proportional to $h(x) \times \mathrm{Prob[jump ~of ~length~} x \in B]$. This means that the chord supported by the first jump is larger than the jump length (see Fig.~\ref{fig3}) and is given by $\int_x^{\infty} dc \, f(c)$. As a consequence, the density of the first jump inside $B$ will be given by
\begin{equation}
 p_{in}(x) = \frac{\displaystyle h(x) \int_x^{\infty} dc \, f(c)} {\displaystyle \int_0^{\infty} dx \, h(x) \int_x^{\infty} dc \, f(c)}.
\end{equation}
We introduce now the ray distribution function: a ray of length $r$ is defined by the distance of a point inside $B$ to the frontier $\partial B$ of $B$. Let $G(r) = \mathrm{Pr\{|P_1 P_2| \leq r: P_1\in B, P_2 \in \partial B \} }$ be the partition function of the rays. Then, $g(r) = dG(r)/dr$ is the corresponding density function~\cite{Mazzolo04}. The ray distribution is related to the chord length distribution by~\cite{Dixmier}
\begin{equation}
 \displaystyle g(r) = \frac{1}{\langle\,c\,\rangle} \int_r^{\infty} dc \, f(c).
\label{def_g(r)}
\end{equation}
The density of the first jump inside $B$ can therefore be written in the more compact form
\begin{equation}
 p_{in}(x) = \frac{\displaystyle h(x) g(x)} {\displaystyle \int_0^{\infty} dx \, h(x) g(x)}.
\end{equation}

\begin{figure}[!h]
\centering
\includegraphics[width=2.in,height=1.7in]{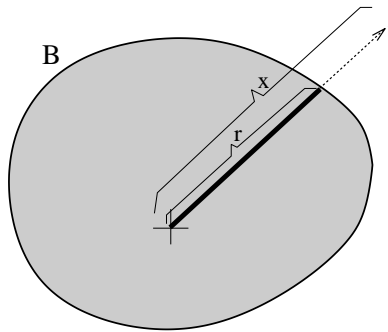}
\setlength{\abovecaptionskip}{15pt} 
\caption{First jumps outside $B$ generate last jumps inside $B$ (formally, a ray).}
\label{fig4}
\end{figure}

The probability density $p_{out}(r)$ of the last jump (this definition excludes the case of chords) can be obtained by following a similar strategy. Actually, $p_{out}(r)$ is proportional to $g(r) \times \mathrm{Prob[jump ~of ~length~} x \not\in B]$, which means that the jump length is larger than the supported ray and is thus given by $\int_r^{\infty} dx \, t(x)$. The function $g(r)$ stems from the collision density being uniform inside the domain, as shown above. By recollecting all these arguments, we are led to 
\begin{equation}
 p_{out}(r) = \frac{\displaystyle g(r) \int_r^{\infty} dx \, t(x)} {\displaystyle \int_0^{\infty} dr g(r) \int_r^{\infty} dx \, t(x)}.
\end{equation}
Dividing both numerator and denominator by $\lambda$, we recognize the density $h(r) = 1/\lambda \int_r^{\infty} dx \, t(x)$ in Eq.~\eqref{eq.h}, and we thus obtain
\begin{equation}
 p_{out}(r) = \frac{\displaystyle h(r) g(r)} {\displaystyle \int_0^{\infty} dr \, h(r) g(r)}.
\end{equation}
It is then apparent that $p_{in}$ and $p_{out}$ are identically distributed. This strong property stems from $\langle \Psi \rangle_\Sigma$ being uniform, which is a consequence of the particular choice of $h(r)$. We therefore drop the subscript \textit{in} and \textit{out}, and name both variables $\tilde{r}$. The mean value $\langle\,\tilde{r}\,\rangle$ is given by
\begin{equation}
\displaystyle \langle\,\tilde{r}\,\rangle = \frac{\displaystyle  \int_0^{\infty} dr \, h(r) g(r)r} {\displaystyle\int_0^{\infty} dr \, h(r) g(r)}.
\label{mean_r_start}
\end{equation}
Replacing Eq.~\eqref{def_g(r)} in the denominator of Eq.~\eqref{mean_r_start} gives 
\begin{align}
 \displaystyle  \int_0^{\infty} dr \, h(r) \, g(r)  & = \displaystyle \int_0^{\infty} dr \, h(r) \frac{1}{\langle\,c\,\rangle} \left[1-\int_0^r dc \, f(c) \right]\notag \\
		& = \frac{1}{\langle\,c\,\rangle} \left[1-\int_0^{\infty} dr \, h(r) \int_0^r dc \, f(c) \right]\notag \\
		& = \frac{1-P_0}{\langle\,c\,\rangle} ,
\end{align}
and thus we get
\begin{equation}
\label{mean_r_final}
 \displaystyle \langle\,\tilde{r}\,\rangle = \displaystyle \frac{\langle\,c\,\rangle}{1-P_0}  \int_0^{\infty} dr \, h(r) g(r)r.
\end{equation}

\subsection{The term $\langle\,\tilde{c}\,\rangle$}

The conditional probability density of sampling a chord at the first jump is proportional to the probability density of chords $f(c)$ (unconstrained chords of length $c$), times the probability that the length of the first jump is larger than the chord, which is $\int_c^{\infty} dr \, h(r)$. The quantity $p_{\tilde{c}}(c)$ is thus given by
\begin{equation}
 \displaystyle p_{\tilde{c}}(c) = \frac{\displaystyle f(c) \int_c^{\infty} dr \, h(r)} {\displaystyle\int_0^{\infty} dc \, f(c) \int_c^{\infty} dr \, h(r)},
\end{equation}
from which follows
\begin{equation}
 \displaystyle \langle\,\tilde{c}\,\rangle = \frac{\displaystyle \int_0^{\infty} dc \, f(c)  \, c \int_c^{\infty} dr \, h(r)} {\displaystyle\int_0^{\infty} dc \, f(c) \int_c^{\infty} dr \, h(r)}.
\label{c_mean_start}
\end{equation}
By observing that the denominator in Eq.~\eqref{c_mean_start} is actually $P_0$, $\langle\,\tilde{c}\,\rangle$ is therefore expressed as
\begin{equation}
 \displaystyle \langle\,\tilde{c}\,\rangle = \frac{1}{P_0} \displaystyle \int_0^{\infty} dc \, f(c) c \int_c^{\infty} dr \, h(r) = \frac{1}{P_0} \displaystyle \int_0^{\infty} dr \, h(r)  \int_0^{r} dc \, f(c) c .
 \label{mean_c_final}
\end{equation}

\begin{figure}[!h]
\centering
\includegraphics[width=2in,height=1.4in]{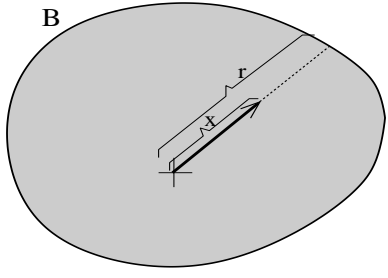}
\setlength{\abovecaptionskip}{15pt} 
\caption{Jumps entirely contained in $B$ must be smaller than the supported ray.}
\label{fig5}
\end{figure}

\subsection{The term $\langle\,\tilde{s}\,\rangle$}

The conditional probability density of having a jump of length $r$ entirely inside $B$ is proportional to the density $t(x)$, times the probability that this jump has a length smaller than the supported ray (see Fig.~\ref{fig5}). This last probability being equal to $\int_{x}^{\infty} dr \,g(r) $, we thus get
\begin{equation}
 p_{\tilde{s}}(x)= \frac{\displaystyle  t(x)  \int_x^{\infty} dr \, g(r)} {\displaystyle \int_0^{\infty} dx \, t(x)  \int_x^{\infty} dr \, g(r)}.
\end{equation}
Hence,
\begin{equation}
 \displaystyle \langle\,\tilde{s}\,\rangle = \frac{\displaystyle \int_0^{\infty} dx \, t(x) x \int_x^{\infty} dr \, g(r)} {\displaystyle \int_0^{\infty} dx \, t(x)  \int_x^{\infty} dr \, g(r)} .
 \label{mean_s_start}
\end{equation}
The integral over $g(r)$ can be easily expressed in terms of the chords density $f(c)$, namely,
\begin{equation}
\label{int_g_r}
 \displaystyle \int_x^{\infty} dr \, g(r) = \int_x^{\infty} dr \, \frac{1}{\langle\,c\,\rangle} \int_r^{\infty} dc \, f(c) = \frac{1}{\langle\,c\,\rangle} \left[ \langle\,c\,\rangle - x + \int_0^x dc \, f(c)(x-c) \right].
\end{equation}
Replacing this identity in the denominator of Eq.~\eqref{mean_s_start} and thanks to Eq.~\eqref{P_0_final}, we get
\begin{equation}
 \displaystyle \int_0^{\infty} dx \, t(x)  \int_x^{\infty} dr \, g(r) = 1-\frac{\lambda}{\langle\,c\,\rangle}(1-P_0).
 \label{denom_mean_s}
\end{equation}
The numerator of Eq.~\eqref{mean_s_start} can be rearranged as follows:
\begin{align}
	\displaystyle \int_0^{\infty} dx \, t(x) \int_0^{x} du \int_x^{\infty} dr \, g(r) & = \lambda - \int_0^{\infty} dx \, t(x) \int_0^{x} du \int_0^{x} dr \, g(r) \notag \\
  & = \lambda - \int_0^{\infty} du \int_u^{\infty} dx \, t(x) \left[\int_0^{\infty} dr \, g(r) - \int_x^{\infty} dr \, g(r) \right]\notag \\
  & = \lambda - \lambda \int_0^{\infty} dr \, h(r) \int_0^{\infty} du \, g(u) - \lambda \int_0^{\infty} dr \int_r^{\infty} du \, g(u) \int_r^{u} dx \, t(x). 
\end{align}
Then, using Eq.~\eqref{eq.h} and the identity $\int_0^{r} du \, g(u) = r g(r) + \frac{1}{\langle\,c\,\rangle} \int_0^{r} dc \, f(c) c$ yields for the numerator
\begin{equation}
\label{num_mean_s}
   \displaystyle \int_0^{\infty} dx \, t(x) \, x \int_x^{\infty} dr \, g(r) = \lambda \left[1- 2 \int_0^{\infty} dr \, g(r) h(r) r - \frac{1}{\langle\,c\,\rangle} \int_0^{\infty} dr \, h(r) \int_0^{r} dc \, f(c) c\right]. 
\end{equation}
Replacing the expression of both numerator and denominator in Eq.~\eqref{mean_s_start} leads to
\begin{equation}
\label{mean_s_final}
 \displaystyle \langle\,\tilde{s}\,\rangle = \frac{\lambda}{1-\frac{\lambda}{\langle\,c\,\rangle}(1-P_0)} \left[1- 2 \int_0^{\infty} dr \, g(r) h(r) r - \frac{1}{\langle\,c\,\rangle} \int_0^{\infty} dr \, h(r) \int_0^{r} dc \, f(c) c\right] .
\end{equation}
Replacing finally Eqs.~\eqref{P_0_final},~\eqref{mean_r_final},~\eqref{mean_c_final} and~\eqref{mean_s_final} in Eq.~\eqref{L_start} gives
\begin{align} 
 \langle\,L\,\rangle_\Sigma & = \displaystyle \langle\,\tilde{c}\,\rangle P_0 + 2 \displaystyle \langle\,\tilde{r}\,\rangle(1- P_0) + \displaystyle \langle\,\tilde{s}\,\rangle \left[\displaystyle \langle\,N\,\rangle -(1-P_0)\right] \notag\\
       & = \frac{1}{P_0} \displaystyle \int_0^{\infty} dr \, h(r)  \int_0^{r} dc \, f(c) c  \, \scriptstyle{\times} \displaystyle \, P_0  \notag \\
       & ~~+ 2 \scriptstyle{\times} \displaystyle \frac{\langle\,c\,\rangle}{1-P_0}  \int_0^{\infty} dr \, h(r) g(r) r \scriptstyle{\times} \displaystyle (1- P_0) \notag \\
       & ~~+ \frac{\lambda}{1-\frac{\lambda}{\langle\,c\,\rangle}(1-P_0)} \left[1- 2 \int_0^{\infty} dr \, g(r) h(r) r - \frac{1}{\langle\,c\,\rangle} \int_0^{\infty} dr \, h(r) \int_0^{r} dc \, f(c) \, c\right] \scriptstyle{\times} \displaystyle \left[\frac{\langle\,c\,\rangle}{\lambda}-(1-P_0)\right]\notag \\
       & = \langle\,c\,\rangle.
\end{align}
Hence follows the central result of this paper: the average length travelled in $B$ by isotropic Pearson random walks with arbitrarily distributed jump lengths satisfies the Cauchy's formula
\begin{equation}
 \langle\,L\,\rangle_\Sigma = \eta_d \frac{V}{\Sigma},
\label{equa_cauchy_L_final} 
\end{equation}
provided that the first jump obeys $h(r)=(1/\lambda) \int_r^{\infty} dx \, t(x)$ and that the walkers enter $B$ uniformly and isotropically. Correspondingly, as shown above, the average number of collisions in $B$ satisfies
\begin{equation}
 \langle\,N \,\rangle_\Sigma = \frac{\eta_d}{\lambda} \frac{V}{\Sigma} =\frac{\langle\,L\,\rangle_\Sigma}{\lambda}.
\label{equa_cauchy_N_final} 
\end{equation}

\section{Absorbing domains}

We conclude our analysis by considering the possibility that the random walks are uniformly absorbed in $B$, which is the case, e.g., for neutrons traversing capturing media. We will focus on purely absorbing domains (the particle trajectories terminate at the first collision), where the probabilistic tools developed in the previous Sections can be straightforwardly applied. In particular, we can show that for this configuration the average number $\langle\,N\,\rangle_\Sigma$ of collisions for trajectories starting on the surface is related to the average number $\langle\,N\,\rangle_{V}$ of collisions for trajectories starting uniformly in the volume. Analogously, a similar relation exists for the travelled lengths.

\begin{figure}[!h]
\centering
\includegraphics[width=4.in,height=1.8in]{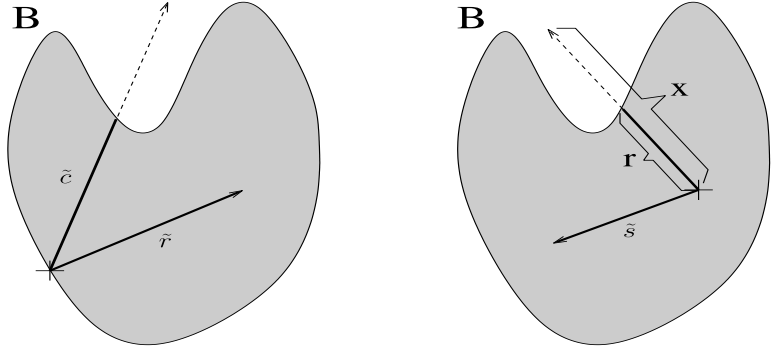}
\setlength{\abovecaptionskip}{15pt} 
\caption{Only jumps falling outside $B$ at the first step contribute to $P_0$ (left) and to $P_{out}$ (right).}
\label{fig6}
\end{figure}

Let us begin from the average number of collisions. We have already introduced $P_0$, namely the probability for a jump starting on the surface to land outside $B$. In the same way, we define $P_{out}$ as the probability for a jump starting in $B$ to land outside $B$. The quantities $\langle\,N\,\rangle_\Sigma$ and $\langle\,N\,\rangle_{V}$ are then respectively given by
\begin{align}
\label{mean_N_start}
	\langle\,N\,\rangle_{\Sigma} & = 1 - P_0 \notag \\
    \langle\,N\,\rangle_{V}      & =   1 -P_{out}. 
\end{align}
The explicit expression for $P_0$ is provided by Eq.~\eqref{P_0_final}, whereas the quantity $P_{out}$ can be determined as follows. Suppose that a walker is isotropically and uniformly emitted in $B$, with jump length density $t(x)$: then, $P_{out}$ corresponds to the probability that the jump length $x$ is larger than the ray of length $r$ (whose density is $g(r)$) in the direction of the jump (see Fig.~\eqref{fig6}). Then, it follows
\begin{equation}
 \displaystyle P_{out} = \int_0^{\infty} dx \, t(x) \int_0^{x} dr \, g(r). 
\label{P_out_start}
\end{equation}
From Eq.~\eqref{int_g_r}, we immediately get
\begin{equation}
 \displaystyle \int_0^{x} dr \, g(r) = \frac{1}{\langle\,c\,\rangle} \left[ x - \int_0^x dc \, f(c)(x-c) \right].
\end{equation}
Replacing this expression in Eq.~\eqref{P_out_start} yields
\begin{align}
	P_{out} & = \frac{1}{\langle\,c\,\rangle} \bigg[ \lambda - \int_0^{\infty} dx \, t(x) \int_0^x dc \, f(c)(x-c) \bigg]       \notag \\
            & =   \frac{\lambda(1 -P_0)}{\langle\,c\,\rangle}.
\label{P_out_final}
\end{align}
By resorting to Eq.~\eqref{mean_N_start}, we finally obtain
\begin{equation}
	\langle\,N\,\rangle_{\Sigma} = \frac{\langle\,c\,\rangle}{\lambda} \Big[ 1 - \langle\,N\,\rangle_{V} \Big].
\label{mean_N_final}
\end{equation}

We consider next the travelled lengths. We define $\langle\,L\,\rangle_{V}$ as the average travelled length when the walkers start uniformly and isotropically in $B$. For purely absorbing media, since trajectories end at the first collision, we have
\begin{align} 
\label{mean_L_start}
	\langle\,L\,\rangle_{\Sigma} & = \langle\,\tilde{c}\,\rangle P_0 + \langle\,\tilde{r}\,\rangle (1 - P_0) \notag \\
     \langle\,L\,\rangle_{V}      & = \langle\,\tilde{r}\,\rangle P_{out} + \langle\,\tilde{s}\,
\rangle (1- P_{out}),
\end{align}
where $\langle\,\tilde{c}\,\rangle,\langle\,\tilde{r}\,\rangle$ and $\langle\,\tilde{s}\,\rangle$ are respectively defined by Eqs.~\eqref{mean_c_final},~\eqref{mean_r_final}, and~\eqref{mean_s_final}. Replacing the explicit expressions for these quantities and those for the probabilities $P_0$ and $P_{out}$ in Eq.~\eqref{mean_L_start} yields
\begin{equation}
 \langle\,L\,\rangle_{\Sigma}  = \int_0^{\infty} dr \, h(r) \int_0^{r} dc \, f(c) c + \langle\,c\,\rangle  \int_0^{\infty} dr \, g(r) h(r) r,
\label{mean_L_sigma_final}
\end{equation}
and
\begin{equation}
 \langle\,L\,\rangle_{V}  = \lambda \left[1- \int_0^{\infty} dr \, g(r) h(r) r - \frac{1}{\langle\,c\,\rangle} \int_0^{\infty} dr \, h(r) \int_0^{r} dc \, f(c) \right].
 \label{mean_L_V_final}
\end{equation}
These equations can be finally combined, and we thus obtain
\begin{equation}
 \langle\,L\,\rangle_{\Sigma} = \langle\,c\,\rangle \Big[ 1 - \frac{1}{\lambda} \langle\,L\,\rangle_{V} \Big].
\label{mean_L_V_S}
\end{equation}
Observe that these formulas, which have been here established for general Pearson walks in purely absorbing media, are surprisingly identical to those previously derived for exponential Pearson walks in the presence of absorption~\cite{Zoia3}. In sharp contrast with the case of diffusive random walks, Eq.~\eqref{mean_L_V_S} depends on the details of the underlying process and of the geometry (because of the $\langle\,L\,\rangle_{V}/\lambda$ term) and has therefore lost its universal character.

\section{Conclusions}

In this paper we have examined under which conditions Cauchy's formulas hold true for diffusive Pearson walks with arbitrarily distributed jump lengths. If the walkers enter the domain $B$ uniformly and isotropically, and the first jump length obeys $h(r) = 1/\lambda \int_r^{\infty} dx \, t(x)$, where $t(x)$ is the probability density of the jump lengths in $B$ and $\lambda = \int_0^{\infty} dx \, x t(x)$ is the average jump size, we have shown that the mean length $\langle L \rangle_{\Sigma} $ travelled by the particles from the surface to the first exit from $B$ (which is proportional to the mean first-return time of the walk) satisfies the Cauchy's formula and is thus universal, in that it does not depend on the features of the random walk. In particular, $\langle L \rangle_{\Sigma} $ is expressed in terms of the ratio between the volume and the surface of $B$ alone, up to a dimension-dependent constant (in three dimensions, we have for instance $\langle L \rangle_{\Sigma} =4 V/\Sigma$). A similar result has been shown to hold also for the average number of collisions performed by the walkers in $B$, which reads $\langle N \rangle_{\Sigma}  = \langle L \rangle_{\Sigma} /\lambda$. 

These findings apply to a broad class of diffusive random walks, and as such extend the range of applicability of Cauchy's formulas, which have been so far established for Pearson walks with exponentially distributed jump lengths. Moreover, the analysis of the case of purely absorbing media seems to suggest that the generalized Cauchy's formulas of~\cite{Zoia3}, derived for exponential Pearson walks, may actually hold true for arbitrary Pearson walks (under the same hypothesis concerning the first jump length density). Investigations are ongoing and will be the subject of further research.

\end{document}